# Modeling the Non-Linear Behavior of Library Cells for an Accurate Static Noise Analysis

Cristiano Forzan
STMicroelectronics, Central R&D
Bologna, 40136 Italy
tel. +39 051 2093829
cristiano.forzan@st.com

Davide Pandini
STMicroelectronics, Central R&D
Agrate Brianza, 20041 Italy
tel. +39 039 6036437
davide.pandini@st.com

## ABSTRACT

In signal integrity analysis, the joint effect of propagated noise through library cells, and of the noise injected on a quiet net by neighboring switching nets through coupling capacitances, must be considered in order to accurately estimate the overall noise impact on design functionality and performances. In this work the impact of the cell non-linearity on the noise glitch waveform is analyzed in detail, and a new macromodel that allows to accurately and efficiently modeling the non-linear effects of the victim driver in noise analysis is presented.

Experimental results demonstrate the effectiveness of our method, and confirm that existing noise analysis approaches based on linear superposition of the propagated and crosstalk-injected noise can be highly inaccurate, thus impairing the sign-off functional verification phase.

## 1. INTRODUCTION

Because of the increasing clock frequencies, the use of high-performance design styles, like domino logic and pass-transistor logic, and the growing importance of the wiring coupling capacitance with respect to the gate and interconnect ground capacitance, the crosstalk-induced noise has become one of the limiting factors of the performances and functionality of large VLSI digital designs. Therefore, Static Noise Analysis (SNA) is a crucial task in the sign-off verification phase of the top-down ASIC design flow [1][2][3]. Basically, noise analysis consists of two different steps: first, the crosstalk-injected noise glitch amplitude and width (or area) are computed for the net under consideration (i.e., the *victim* net); then, the total noise glitch is obtained with the combination of the noise injected by the neighboring *aggressor* nets, and the noise propagated through the victim driver cell. In the second step, the noise at the victim receiver is compared against dynamic noise margins [4], represented by the *Noise Rejection Curve* (NRC). When the noise waveform width (or area) and amplitude are in the NRC failure region (i.e., above the curve), the noise analysis tool flags an error.

In several approaches, either the propagated noise is not considered at all, or the propagated and injected noise are evaluated separately. The noise propagating from the input to the output of the victim driver cell is usually obtained from pre-characterized tables as a function of the input noise glitch area (or width) and height. In contrast, the noise injected on a victim net by adjacent aggressor nets through coupling capacitances is typically computed with network Model Order Reduction (MOR) techniques [3][5], or analytical macromodels that capture the relevant physical parameters of the noise waveform [6], and by assuming a linear model of both the aggressor and the victim driver cells (the Thevenin-equivalent circuit for the aggressor drivers, and the holding resistance for the victim driver). The linearity assumption allows using superposition thus simplifying the worst-case identification that occurs when all the noise glitch peaks are aligned. However, in digital systems the library cells have a strong non-linear behavior, and a simple summation of the two noises may lead to a large underestimation of the total noise, thus potentially leaving many functional failures undetected.

This problem was investigated in [4], where Zolotov *et al.* proposed an iterative approach to represent the victim driver with a Thevenin-equivalent circuit consisting of a pulsed voltage source and a resistance, in order to capture the non-linear behavior of the driver with a linear model, and carry out a noise analysis using MOR techniques and linear superposition. However, their approach may still yield large errors in both the noise peak (-18%) and width (-20%). Such large underestimations in the total noise glitch would leave many potential functional failures undetected in SNA.

This paper is organized as follows: in Section 2 our macromodel for modeling the victim driver non-linear behavior is presented, and experimental results showing the effectiveness of our method are reported and discussed in Section 3. Finally, Section 4 summarizes some conclusive remarks.

## 2. PROPOSED MACROMODEL

In our approach, the victim driver is modeled with a Voltage-Controlled Current Source (VCCS) $I_{DC}$, similarly to [4]. This current source is a non-linear function of both the cell input and output voltages, and is expressed as:

$$I_{DC} = f(V_{in}, V_{out}), \qquad (1)$$

which is obtained during a pre-characterization step, by performing a simple DC analysis, where $V_{in}$ and $V_{out}$ are swept across the characterization range corresponding to the typical voltage swing of the given technology.

A distributed RC network including the coupling capacitances represents the interconnections within the noise cluster[1], the aggressor driver linear models are Thevenin equivalent circuits (where $V_{TH}$ is a saturated ramp and $R_{TH}$ is the driving resistance) obtained as in [7], and the victim and aggressor receivers can be modeled by their input capacitances. Since the total noise waveform must be accurately evaluated at the victim driver output (or the victim net driving point), the coupled interconnect network is modeled at the driving points. Therefore, the driving point impedance of the interconnect network is represented by a coupled-$\pi$ model, which can be obtained with moment-matching techniques following the approach presented in [8]. The corresponding noise cluster macromodel of a victim and two coupled aggressors is shown in Figure 1.

---

[1] A victim net and its neighboring coupled aggressors are referred as *noise cluster*, or simply as *cluster*.

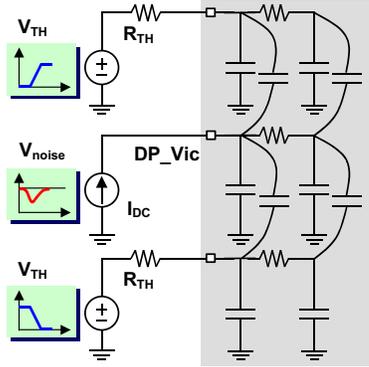

**Figure 1. Noise cluster macromodel**

Although this macromodel contains the non-linear VCCS $I_{DC}$ represented by the load curves (1), linear analysis and superposition can still be exploited by using the VCCS values computed during the pre-characterization step. Moreover, since the noise cluster macromodel is a simple circuit, the total noise waveform can be accurately and efficiently computed by means of a dedicated engine embedded into the noise analysis tool. Our approach can be straightforwardly extended to clusters with several aggressors with different switching directions and phase alignments.

## 3. EXPERIMENTAL RESULTS

In order to highlight the non-linear behavior of the victim driver cell, a simple test case in 0.13µm technology, consisting of two adjacent coupled nets, was considered. The wiring parasitics were extracted from two 500µm parallel-running interconnects, designed on metal layer 4, where the aggressor cell is an inverter and the victim driver is a 2-input nand. The results from circuit simulations performed with ELDO™ [9] and reported in Table 1 clearly demonstrate that linear superposition between injected and propagated noise may induce a significant underestimation of the total noise. As a consequence, SNA based on NRCs can be quite inaccurate and may fail to report several noise-induced functional failures.

**Table 1. Injected and propagated noise combination**

| Noise | ELDO™ | Linear superposition | Error% | Our macromodel | Error% |
|---|---|---|---|---|---|
| Peak (V) | 0.345 | 0.269 | -22.0 | 0.354 | 2.6 |
| Area (V·ps) | 174.3 | 82.18 | -52.8 | 175.7 | 0.8 |

Table 2 shows the numerical results of two in-phase aggressors and one propagating noise glitch through the victim driver (2-input nand).

**Table 2. Worst-case overlapping between two aggressors and one propagating noise glitch**

| Noise | ELDO™ | Our macromodel | Error% |
|---|---|---|---|
| Peak (V) | 0.919 | 0.947 | 3.1 |
| Area (V·ps) | 496.2 | 508.7 | 2.5 |

Our approach has been tested on several noise clusters in 0.13µm and 90nm technology, and its accuracy evaluated against circuit simulations, and the error was always within few percents. The speed-up obtained with our approach was about 20X with respect to ELDO™, thus yielding a practical approach for noise analysis.

## 4. CONCLUSIONS

In this work we have proposed a macromodel for the victim driver non-linearity and generates a very accurate representation of the total noise glitch waveform at the victim driver output, when both propagated and crosstalk-injected noise are present. Our approach has been validated on a broad range of victim-aggressor configurations, and the experimental results show an excellent accuracy with respect to circuit simulations, with a significant computational speed-up. Future work will focus on developing a complete methodology for static noise analysis based on our macromodel.